\documentclass{ifacconf}
\usepackage{comment}
\usepackage{times,mathptmx}
\usepackage{color}
\usepackage{epsfig}
\usepackage{rotating}
\usepackage{amsmath}
\usepackage{amssymb}
\usepackage{amsfonts}
\usepackage{url}
\usepackage{tikz}
\usepackage{pgfplots}
\usepackage{import}
\usepackage{algorithm,mathtools}
\usepackage{comment}
\usepackage[noend]{algpseudocode}
\usepackage{epsf}
\usepackage{psfrag}
\usepackage{graphicx}      
\usepackage{natbib}        

\usetikzlibrary{shapes,arrows}
\begin{document}
\begin{frontmatter}

   \title{Learning Models of Model Predictive Controllers using Gradient Data \thanksref{footnoteinfo}}
   \thanks[footnoteinfo]{This work was partially supported by the Swedish Research Council and by the Wallenberg AI, Autonomous Systems and Software Program (WASP) funded by the Knut and Alice Wallenberg Foundation and the Swedish Research Council Research Environment NewLEADS under contract 2016-06079.}
   
   \author[First]{Rebecka Winqvist} 
   \author[First]{Arun Venkitaraman} 
   \author[First]{Bo Wahlberg}
   
   \address[First]{Division of Decision and
           Control Systems, EECS, KTH Royal Institute of Technology, Stockholm, Sweden. (e-mails: \{rebwin, arunv, bo\}@kth.se).}

   \begin{abstract}                
   This paper investigates controller identification given data from  a Model Predictive Controller (MPC) with constraints. 
   We propose an approach for learning MPC that explicitly uses the gradient information in the training
   process. This is motivated by the observation that recent differentiable convex optimization MPC solvers
   can provide both the optimal feedback law from the state to control input as well as the corresponding 
   gradient. As a proof of concept, we apply this approach to explicit MPC (eMPC), for which the feedback 
   law is a piece-wise affine function of the state, but the number of pieces grows rapidly with the 
   state dimension.  Controller  identification can here be used to find an approximate lower complexity
   functional approximation of the controller. The eMPC is modelled with a Neural Network (NN)
   with Rectified Linear Units (ReLUs), since such NN can represent any  piece-wise affine function. 
   A motivation is to replace on-line solvers with neural networks to implement  MPC and to simplify 
   the evaluation of the function in larger input dimensions. We also 
   study experimental design and  model evaluation in this framework, and propose a hit and run sampling
   algorithm for input design. The proposed algorithm are illustrated and numerically evaluated on a second order MPC problem.
   
   \end{abstract}

\end{frontmatter}


\section{Introduction}
\label{Sec1}
Controller identification concerns estimating a model of a feedback controller
from observed input and output data. This is typically done in a feedback mode, 
where the controller interacts with a dynamical system. It is a well studied topic, 
see e.g.~\citep{ljung1999system}, in particular when the system and the controller 
can be described by linear dynamical models. It is only recently that learning models 
based on Neural Networks (NNs) have been pursued for Model Predictive Controllers (MPCs) 
with constraints \citep{2018_paper,invariant_sets,chen2019large,maddalena2019neural}. 
A motivation behind these approaches is the recent result that NNs with rectified linear 
units (ReLUs) can represent piece-wise affine functions or functions with linear regions 
\citep{NIPS_ReLU}. This naturally makes such approaches suited to the MPC learning problem 
\citep{2018_paper}, particularly in the case of explicit MPC, where the optimal control law 
is an affine function of the state vector \citep{empc_bemporad}.

Consider an MPC problem with the feedback law $\mathbf{u}=\mathbf{u}^*(\mathbf{x})$ 
corresponding to the full state information $\mathbf{x}$. The goal of any learning 
based approach to MPC is then to learn a mapping $\mu(\mathbf{x})$ that describes
$\mathbf{u}^*(\mathbf{x})$ as good as possible.  In the case of NN based approaches, 
$\mu$ corresponds to the function learnt by a NN.  Most existing NN based learning 
approaches proceed by learning $\mu$ using only the input and output observations 
of $\mathbf{x}$ and $\mathbf{u}$, respectively. Like any learning approach, an MPC 
mapping learnt in such a manner could perform poorly when the training data is limited. 
In such a scenario, additional structure can often be of merit in aiding the training of 
the NN. For example, in the case of explicit MPC  (eMPC), $\mathbf{u}$ is a piece-wise 
affine function in $\mathbf{x}$ and hence, the gradient of the optimal control law with 
respect to $\mathbf{x}$ is piecewise-constant and contains important structural information 
useful in learning $\mu$. 

Motivated by this observation, we propose a NN-based learning approach for MPC where we 
explicitly use structural information in the form of gradient data 
$\frac{\partial\mathbf{u}^*}{\partial\mathbf{x}}$ for training the NN. While training data 
in the form of gradient information is typically unavailable or difficult to obtain, the special 
structure of the MPC problem and the use of recently proposed tools in differentiable convex 
optimization \citep{CVXPY} helps us achieve our goal. 
The main contributions of the paper are:
\begin{itemize}
\item {\it Learning Models}:  We design and evaluate algorithms to train ReLU 
                              based NNs that implements MPC using  input and output 
                              data $\mathbf{u}_i=\mathbf{u}^*(\mathbf{x}_i)$) and 
                              corresponding gradient data $\mathbf{u}_i'=\partial\mathbf{u}^*(\mathbf{x})/\partial\mathbf{x}| \mathbf{x}=\mathbf{x}_i$. 
                              We show that taking the gradient information into 
                              account can significantly reduce the number of training 
                              data needed to achieve a high accuracy. We use eMPC as a 
                              proof of concept, while the proposed algorithm can handle 
                              more general MPC problems.

\item {\it Data generation}: It is not obvious how to efficiently generate training data 
                             when learning the control law as a mapping. Grid-based approaches 
                             for sampling the input space would work for small state dimensions, 
                             but become cumbersome for high-dimensional systems. Keeping this in 
                             mind, we study the use of an efficient and statistically motivated 
                             hit-and-run sampler that extends well to higher state dimensions.

\item {\it Evaluation of performance}: The performance of trained NN based MPC controller 
                                       should be evaluated in closed loop feedback, taking into 
                                       account both tracking, disturbance rejection and stability. 
                                       Here, we evaluate the performance on test data in terms of different metrics. 

\end{itemize}

\subsection{A motivating example: Identification using Gradient Data}
In order to illustrate the identification algorithms to be proposed, consider the following scalar 
linear feedback example with measurements 
\begin{equation*}
 u_k=l_1x_k+l_2+e_k, \quad k=1,\ldots, N,
\end{equation*}
with control signal $u_k$, scalar state signal $x_k$ and additive white zero mean 
Gaussian noise $e_k$ with variance $\lambda_e$. Assume that it is possible to measure 
the derivative of $u$ with respect to~$x$,
\begin{eqnarray*}
u'_k&=l_1 +v_k, \quad k=1,\ldots, N,
\end{eqnarray*}
where $v_k$ is white zero mean Gaussian distributed noise with variances $\lambda_v$.
The maximum likelihood estimate of $l_1$ and $l_2$ given the data $\{x_k,u_k,u'_k\}$, $k=1,\ldots, N$ 
is found by solving the least squares problem
\begin{equation*}
V(l_1, l_2)= \sum _{k=1}^N \frac{[u_k-l_1x_k-l_2]^2}{\lambda_e}+ \frac{[u'_k-l_1]^2}{\lambda_v}.
\end{equation*}
The error covariance matrix of the least squares estimate $(\hat{l}_1,\; \hat{l}_2)^T$ equals, 
see  \citet{ljung1999system},
\begin{equation*}
\frac{\lambda_e}{N} \left[ \frac{1}{N}\sum_{k=1}^{N}\begin{bmatrix}
(x_k^2 +\lambda_e/\lambda_v) & x_k\\  x_k & 1
\end{bmatrix}\right]^{-1}
\end{equation*}
For the choice $x_k=1$ it is not possible to estimate $l_1$ and $l_2$ individually 
without the extra gradient data. For this case the estimation error covariance matrix equals
\begin{equation*}
\frac{1}{N} \begin{bmatrix}
\lambda_v & -\lambda_v\\  -\lambda_v & (\lambda_e+\lambda_v)
\end{bmatrix}
\end{equation*}
This result can also be found by analyzing
\begin{equation*}
u_k-u'_k=b_2+e_k-v_k.
\end{equation*}
The least squares estimate of $l_2$ is just the average of this difference signal. 
Notice that the variance of $\hat{l}_1$ is lower than the variance of  $\hat{l}_2$, 
which is expected given the extra information on ${l}_1$ and the added noises 
when estimating ${l}_2$. Hence gradient information can be crucial in terms of 
estimation quality for low input excitation.

The structure of this paper is as follows: The MPC problem and Neural Networks are 
introduced in Section \ref{sec:preliminaries}, while Section \ref{sec:approach} describes 
the proposed training and evaluation  framework. Data generation and testing are 
studied in Section \ref{sec:data} and the numerical examples are presented in 
Section \ref{sec:examples}. Finally, the conclusion and ideas for future work are given 
in Section \ref{sec:conclusion}.

\section{Preliminaries}
\label{sec:preliminaries}
In this section, we briefly review the basic Model Predictive Control (MPC) problem, 
the explicit MPC, followed by a review of neural networks.

\subsection{Model Predictive Control}
Consider a discrete-time linear time-invariant system which evolves in time as
\begin{equation}
    \mathbf{x}({k+1}) = \mathbf{Ax}(k) + \mathbf{Bu}(k),
    \label{eq:system}
\end{equation}
where $\mathbf{x}(k)$ denotes the state vector and $\mathbf{u}(k)$ denotes the input or 
control action, at the $k$th time-step,respectively; $\mathbf{A}$ and $\mathbf{B}$ denote 
the system matrices. Model-predictive control (MPC) refers to the problem of
steering the state of the system \eqref{eq:system} from an initial value to the origin by 
minimizing a control objective, subject to the state and input constraints
\begin{equation}
    \mathbf{x}(k) \in \mathcal{X}, \quad \mathbf{u}(k) \in \mathcal{U}
    \label{eq:constraints}
\end{equation}
where $\mathcal{X} \subseteq \mathbb{R}^n$ and $\mathcal{U} \subseteq \mathbb{R}^m$ are 
polyhedra representing the constraint sets for the state and the input, respectively.
We consider the following fixed horizon MPC problem, see \citet{mpc_book}:
\begin{equation}
    \begin{aligned}
        & \min_{\mathbf{u}(0),\ldots, \mathbf{u}({N-1})}& &  \mathbf{x}(N)^T\mathbf{Q}_N\mathbf{x}(N) + \sum_{k=0}^{N-1}\mathbf{x}(k)^T\mathbf{Q}\mathbf{x}(k)
        + \mathbf{u}(k)^T\mathbf{Ru}(k)
        \\
        & \text{s.t.}
        & &\mathbf{x}(k+1) = \mathbf{Ax}(k) + \mathbf{Bu}(k), \\
        & {} & & \mathbf{x}(k) \in \mathcal{X}, \\
        & {} & & \mathbf{u}(k) \in \mathcal{U}, \\
        & {} & & \mathbf{x}(0) = {\mathbf{x}}
    \end{aligned}
    \label{eq:mpc}
\end{equation}
where $\mathbf{Q}$ and $\mathbf{R}$ are the positive semi-definite weight matrices, $\mathbf{Q}_N$ 
the terminal cost matrix, and $N$ denotes the finite time horizon length. The optimal solution to 
such a constrained quadratic program can be found by solving a set of linear equations once 
the active constraints are identified, see \citet{5153127} for results on fast 
online MPC implementations. 

The MPC control law is the mapping from the current state $\mathbf{x}(0) = {\mathbf{x}}$ to 
the first optimal control action $\mathbf{u}(0)=\mathbf{u}$. We will use the notation
$
\mathbf{u}=\mathbf{u}^*(\mathbf{x})
$
for this mapping.

\subsection{Feasibility}
We now discuss how the feasibility constraints can be characterized in terms of set-invariance, 
which will form the basis of incorporating structure into neural network solutions for MPC. 
As defined by \citet{mpc_book}, a set, $\mathcal{C} \subseteq \mathcal{X}$, is a control invariant 
set for the system (\ref{eq:system}) subject to the constraints (\ref{eq:constraints}) if
\begin{equation}
\mathbf{x}(k) \in \mathcal{C} \implies \exists \mathbf{u}(k) \text{ s.t. } \mathbf{x}({k+1}) \in \mathcal{C}, \quad k=1,2,\cdots
\end{equation}
In other words, for any initial state in $\mathcal{C}$ there exists a controller that ensures 
all future states reside in $\mathcal{C}$. The \textit{maximal control invariant} set, $\mathcal{C}_{\infty}$, 
is then defined as the control invariant set containing all control invariant sets contained 
in $\mathcal{X}$. $\mathcal{C}_\infty$ being a polytope is expressible as an intersection
of halfspaces as $\mathcal{C}_{\infty}  \{\mathbf{x} \in \mathbb{R}^n \mid \mathbf{C}_x\mathbf{x} \leq \mathbf{d}_x\}$ \citep{empc_bemporad}. 
To compute $\mathcal{C}_{\infty}$ we use algorithm 10.2 \textit{'Computation of $\mathcal{C}_{\infty}$'} 
provided in \citet{mpc_book} and the accompanying software. 

\subsection{Explicit MPC}
The main challenge in solving the MPC problem lies in determining which of the constraints are active. 
Nevertheless, this entails solving an optimization problem at each time instant, which could quickly 
become a bottleneck when applying to systems repeatedly. One of the ways of circumventing the 
determination of active sets is through offline pre-computation of the control laws such that the 
problem is transformed into that of specifying a mapping or lookup table in the form of a piece-wise 
affine function. This mapping then acts on the input to produce the optimal control law. This approach 
is known as the explicit MPC \citep{Alessio2009,10.5555/1965221,empc_bemporad}. 

Given a polytopic set $\mathcal{X}$, for each $\mathbf{x} \in \mathcal{X}$, explicit MPC computes 
a piecewise affine (PWA) mapping $\mathbf{u}=\mathbf{u}^*(\mathbf{x})$ from $\mathbf{x}$ to 
$\mathbf{u}$ defined over $M$ regions of $\mathcal{X}$. This in turn means that the gradient 
$\partial\mathbf{u}^*(\mathbf{x})/{\partial\mathbf{x}}$ is piecewise-constant and contains 
significant  information describing the optimal control law. This structural information can 
in turn be used to improve the training of the NNs - giving the gradient $\mathbf{u}'$ in 
addition to the state vector $\mathbf{x}$ and the associated with the control law $\mathbf{u}$ 
will help learn a better $\mu(\mathbf{x})$, particularly when the training data is limited in size.

The explicit MPC has further inspired the viewing of the MPC problem as a general 
learning problem. As a result, a number of works involving the use of learning approaches, 
primarily in the form of artificial neural networks have been proposed recently 
\citep{2018_paper,invariant_sets,chen2019large,maddalena2019neural}. The idea of using neural 
networks based MPC is by no means novel and there exist many publications on this topic, 
\citep{Parisini1995ARR,akesson2006,Winqvist1477561}. 

\subsection{Neural networks}
Neural networks and deep learning approaches are now ubiquitous and form the crux of most 
learning-based approaches today \citep{deeplearning}. Neural networks learn a mapping from 
the input to the output from known training examples, when the problem at hand has either 
no clear closed-form input-output mapping, or even if there is one, is intractable to work 
with \citep{Bishop}. Neural networks comprise concatenated processing units known as neurons 
that combine linear and non-linear transformations. Mathematically expressed, a neural network 
$\mu(\mathbf{x})$ learns a mapping from the input $\mathbf{x}$ to output $\mathbf{u}$ in the form
\begin{equation*}
    \mathbf{u}=\mu(\mathbf{x},\pmb\theta)=\sigma(\mathbf{W}_L \sigma(\mathbf{W}_{L-1}\sigma(\cdots \sigma(\mathbf{W}_1\mathbf{x}+\mathbf{b}_1)\cdots)+\mathbf{b}_L),
\end{equation*}
where $\sigma(\cdot)$ denotes the point-wise nonlinearity or activation function, and the 
matrices $\{\mathbf{W}_i,\mathbf{b}_i\}_{i=1}^L$ are the parameters $\pmb\theta$ (weights and biases) 
learnt by the network from the training data, $L$ denoting the number of neuron layers. 

The learning is typically performed by the use of back-propogation that uses the 
gradients of an error or loss function with respect to the network parameters. 
For the reasons of computational complexity and stability, the rectified linear unit 
(ReLU) is the most commonly employed activation function \citep{Bishop,deeplearning}. 
As discussed earlier, the use of ReLU as the activation function has been shown to be 
well motivated in learning functions with linear regions \citep{NIPS_ReLU}, and particularly 
in the MPC setting due to its piece-wise linear nature \citep{2018_paper}. A schematic of 
a two-layer neural network is shown in Figure \ref{fig:noprojNN}.

\def\layersep{1cm}
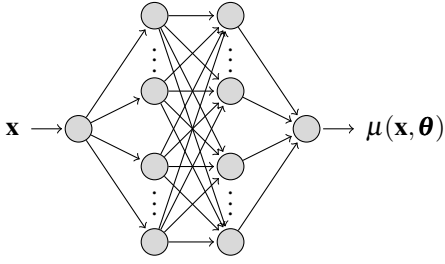
\begin{figure}[t]
\centering
\begin{tikzpicture}[shorten >=1pt,->,draw=black, node distance=\layersep]
    \tikzstyle{every pin edge}=[<-,shorten <=1pt]
    \tikzstyle{neuron}=[draw,circle,fill=blue!60,minimum size=10pt,inner sep=0pt]
    \tikzstyle{input neuron}=[neuron, fill=gray!30];
    \tikzstyle{output neuron}=[neuron, fill=gray!30];
    \tikzstyle{hidden neuron}=[neuron, fill=gray!30];
    \tikzstyle{annot} = [text width=4em, text centered]
    \tikzstyle{missing} = [
    draw=none, 
    scale=1,
    text height=0.1cm,
    execute at begin node=\color{black}$\vdots$
  ]
  \tikzstyle{projection layer}=[draw, thick, rectangle, minimum height = 4em,
    minimum width = 3em, text centered, text width = 3.5em]
    \tikzstyle{fill neuron}=[circle,fill=white,minimum size=2pt,inner sep=0pt]

    \node[input neuron, pin=left:$\mathbf{x}$] (I-1) at (0,-2) {};

    \path[yshift=0.5cm]
    node[hidden neuron] (H-1) at (\layersep,-1 cm) {};
    \path[yshift=0.5cm]
    node[missing] (m3) at (\layersep,-1.7 cm) {};
    \path[yshift=0.5cm]
    node[hidden neuron] (H-2) at (\layersep,-2 cm) {};
    \path[yshift=0.5cm]
    node[hidden neuron] (H-3) at (\layersep,-3 cm) {};
    \path[yshift=0.5cm]
    node[missing] (m4) at (\layersep,-3.6 cm) {};
    \path[yshift=0.5cm]
    node[hidden neuron] (H-4) at (\layersep,-4 cm) {};
    
    \path[yshift=0.5cm]
    node[hidden neuron] (H2-1) at (2*\layersep,-1 cm) {};
    \path[yshift=0.5cm]
    node[missing] (m5) at (2*\layersep,-1.7 cm) {};
    \path[yshift=0.5cm]
    node[hidden neuron] (H2-2) at (2*\layersep,-2 cm) {};
    \path[yshift=0.5cm]
    node[hidden neuron] (H2-3) at (2*\layersep,-3 cm) {};
    \path[yshift=0.5cm]
    node[missing] (m6) at (2*\layersep,-3.6 cm) {};
    \path[yshift=0.5cm]
    node[hidden neuron] (H2-4) at (2*\layersep,-4 cm) {};

    \node[output neuron, pin={[pin edge={->}]right:$\mu(\mathbf{x}, \pmb{\theta})$}] (O2) at (3*\layersep,-2 cm) {};
    
    \foreach \source in {1,...,1}
        \foreach \dest in {1,...,4}
            \path (I-\source) edge (H-\dest);
            
    \foreach \source in {1,...,4}
        \foreach \dest in {1,...,4}
            \path (H-\source) edge (H2-\dest);
    \foreach \source in {1,...,4}
        \path (H2-\source) edge (O2);

\end{tikzpicture}
\caption{Schematic of a two-layer neural network.
}
\label{fig:noprojNN}
\end{figure}


\section{Proposed Approach}
\label{sec:approach}
Let $\mathbf{x}\in\mathbb{R}^{n_x}$  denote the initial state vector $\mathbf{x}(0)$, 
and $\mathbf{u}\in\mathbb{R}^{n_u}$ the corresponding optimal control law $\mathbf{x}(0)$. 
Further, let $\mathbf{u}'\in\mathbb{R}^{n_u\times n_x}$ denote the true gradient of the 
optimal control with respect to $\mathbf{x}$. Consider that we are given a set of $N_{tr}$ 
samples of triplets of initial states, the corresponding optimal control laws, and their 
gradients, given by $\{\mathbf{x}_i,\mathbf{u}_i, \mathbf{u}'_i\}_{i=1}^{N_{tr}}$. We 
note that the subscript $i$ here denotes the $i$th samples and is not the time-index. 
Let us define the sets  $\mathcal{X}=\{\mathbf{x}_1,\cdots,\mathbf{x}_{N_{tr}}\}$, 
$\mathcal{U}=\{\mathbf{u}_1,\cdots,\mathbf{u}_{N_{tr}}\}$, 
and $\mathcal{U}'=\{\mathbf{u}'_1,\cdots,\mathbf{u}'_{N_{tr}}\}$. Our goal is to train a 
ReLU NN to predict the optimal control law using gradient information -- we learn the 
mapping $$\mu(\mathbf{x},\pmb\theta): \mathbb{R}^{n_x}\times\mathbb{R}^D\mapsto\mathbb{R}^{n_u},$$
where $\pmb\theta\in\mathbb{R}^D$ denotes the all the learnable weights of the NN, 
by using a training loss function $\mathcal{L}_{tr}(\pmb\theta): \mathcal{X}\times\mathcal{U}\times\mathcal{U}'\mapsto\mathbb{R}$ 
that explicitly uses gradient information. 

Specifically, we propose to learn $\mu(\mathbf{x},\pmb\theta)$ by minimizing the following 
loss function with respect to the parameters $\pmb\theta$:
\begin{align}
\label{eq:train_loss}
\mathcal{L}_{tr}(\pmb\theta)= \sum_{i=1}^{N_{tr}}\left(\|\mathbf{u}_i-\mu(\mathbf{x}_i,\pmb\theta)\|_2^2+\gamma\left\|\mathbf{u}_i'-\frac{\partial\mu(\mathbf{x},\pmb\theta)}{\partial\mathbf{x}}_{|\mathbf{x}=\mathbf{x}_i}\right\|_2^2\right),
\end{align}
where the first term denotes the model error, and the second term denotes the fit of 
the gradients , $\gamma>0$ being the regularization constant. We note that typically 
NN training is performed by minimizing only the model error $$\sum_{i=1}^{N_{tr}}\|\mathbf{u}_i-\mu(\mathbf{x}_i,\pmb\theta)\|_2^2.$$
It is well known that training the NN in such a manner makes it data-hungry, and the 
performance of the NN often suffers when the number of training samples is limited.

In many control problems, and particularly with the MPC, often one has access only to 
a limited  number of training observations given the very large state space.
As seen from our discussion in Section 2, an explicit incorporation of structural 
information often aids in the learning process when the number of training samples is 
limited. This forms the motivation behind our approach - the regularization explicitly 
enforces structural information on the learnt NN mapping $\mu$. In the case of eMPC, 
the set feasible inputs is a polytope, and therefore, the gradient of the optimal 
control law $\mathbf{u}'$ is piecewise constant. Assuming the training samples are 
drawn randomly over the feasible input space (ideally one sample per region of the 
feasible space), if the samples are taken uniformly from the feasible set, the 
network is given information about a large subsets of the feasible set. In such a 
case, we would expect that our approach would learn with even limited training data 
due to active incorporation of this structural information. In contrast, a regular 
NN based MPC solver that uses only the model error in the training is agnostic to 
this information and must abstract such structure purely from the training samples. 

Thus, our approach is a trade-off between completely data-driven and structurally 
aware MPC solver. A schematic of the proposed approach is given in Figure 
\ref{fig:noprojNN}. We note that our approach requires the value of the true gradients 
of the MPC problem evaluated at the given $\mathbf{x}$. As described in Section 5, 
we evaluate the true gradients using the recently proposed {\em cvxpylayers} 
\citep{agrawal2019differentiable}, a framework for obtaining differentiable convex 
layers in pytorch. Since the cost function $\mathcal{L}_{tr}(\pmb\theta)$ is non-convex, 
the training proceeds by backpropogation as described next. The details of the dataset 
generation and training are described in the next Section. Let $\hat{\pmb\theta}$ denote 
the NN parameters obtained after training. Once the network parameters $\pmb\theta$ are 
learnt, they are used to predict the optimal control law for test data $\mathbf{x}$ as 
$$\hat{\mathbf{u}}=\mu(\mathbf{x},\hat{\pmb\theta}).$$
We note that the gradient information is {\emph{not}} required during the test phase, 
and is used only in the training of the NN.

\def\layersep{1cm}
\begin{figure}[h]
\centering
\begin{tikzpicture}[shorten >=1pt,->,draw=black, node distance=\layersep]
    \tikzstyle{every pin edge}=[<-,shorten <=1pt]
    \tikzstyle{neuron}=[draw,circle,fill=blue!60,minimum size=10pt,inner sep=0pt]
    \tikzstyle{input neuron}=[neuron, fill=gray!30];
    \tikzstyle{output neuron}=[neuron, fill=gray!30];
    \tikzstyle{hidden neuron}=[neuron, fill=gray!30];
    \tikzstyle{annot} = [text width=4em, text centered]
    \tikzstyle{grad block} = [draw, rectangle, minimum size=10pt, fill=gray!30]
    \tikzstyle{missing} = [
    draw=none, 
    scale=1,
    text height=0.1cm,
    execute at begin node=\color{black}$\vdots$
  ]
  \tikzstyle{projection layer}=[draw, thick, rectangle, minimum height = 4em,
    minimum width = 3em, text centered, text width = 3.5em]
    \tikzstyle{fill neuron}=[circle,fill=white,minimum size=2pt,inner sep=0pt]

    \node[input neuron, pin=left:{\color{black}$\mathbf{x}$}] (I-1) at (0,-2) {};

    \path[yshift=0.5cm]
    node[hidden neuron] (H-1) at (\layersep,-1 cm) {};
    \path[yshift=0.5cm]
    node[missing] (m3) at (\layersep,-1.7 cm) {};
    \path[yshift=0.5cm]
    node[hidden neuron] (H-2) at (\layersep,-2 cm) {};
    \path[yshift=0.5cm]
    node[hidden neuron] (H-3) at (\layersep,-3 cm) {};
    \path[yshift=0.5cm]
    node[missing] (m4) at (\layersep,-3.6 cm) {};
    \path[yshift=0.5cm]
    node[hidden neuron] (H-4) at (\layersep,-4 cm) {};
    
    \path[yshift=0.5cm]
    node[hidden neuron] (H2-1) at (2*\layersep,-1 cm) {};
    \path[yshift=0.5cm]
    node[missing] (m5) at (2*\layersep,-1.7 cm) {};
    \path[yshift=0.5cm]
    node[hidden neuron] (H2-2) at (2*\layersep,-2 cm) {};
    \path[yshift=0.5cm]
    node[hidden neuron] (H2-3) at (2*\layersep,-3 cm) {};
    \path[yshift=0.5cm]
    node[missing] (m6) at (2*\layersep,-3.6 cm) {};
    \path[yshift=0.5cm]
    node[hidden neuron] (H2-4) at (2*\layersep,-4 cm) {};

    \node[output neuron, pin={[pin edge={->}]right:{\color{black}$\mu(\mathbf{x}, \pmb{\theta})$}}] (O2) at (3*\layersep,-2 cm) {};
    \node[grad block, pin={[pin edge={->}]right:{\color{black}$\mu'(\mathbf{x}, \pmb{\theta})$}}] (grad) at (3*\layersep,-3.5 cm) {$\frac{\partial \mu }{\partial \mathbf{x}}$};
    \path (O2) edge (grad);
    
    \foreach \source in {1,...,1}
        \foreach \dest in {1,...,4}
            \path (I-\source) edge (H-\dest);
            
    \foreach \source in {1,...,4}
        \foreach \dest in {1,...,4}
            \path (H-\source) edge (H2-\dest);
    \foreach \source in {1,...,4}
        \path (H2-\source) edge (O2);

\end{tikzpicture}
\caption{The structure of the proposed neural network.}
\label{fig:noprojNN}
\end{figure}
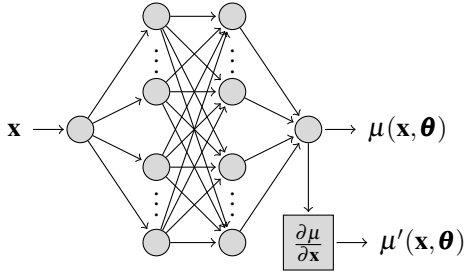


\section{Training and data generation}
\label{sec:data}
We first discuss the systematic strategy that we propose for the generation of 
data sets for the MPC problems. Our approach for generating the training and 
testing data set involves sampling (using the Hit-and-run sampler described next) 
a set of states $\mathcal{X} = \{\mathbf{x}_1, \hdots, \mathbf{x}_{N_{tr}}\}$ from 
the feasible region $\mathcal{C}_{\infty}$ described in Section 3.3. The corresponding 
optimal control input and the corresponding gradient sets, $\mathcal{U}$, 
and $\mathcal{U}'$ are then computed using a stand-alone cvxpylayers. We follow the 
same strategy to generate test data set with the difference that it does not contain 
gradient data.

\subsection{Sampling}
To sample from the maximal control invariant set $\mathcal{C}_{\infty}$ defined in 
Section 2.2 we use the Hit-and-Run sampler, which is a Markov chain Monte Carlo 
method for sampling uniformly from convex shapes \citep{METE20126}.  Essentially, 
starting from any point in the convex set, the {\color{black} sampler} generates a 
set of points, $\mathcal{S}$, by walking steps of length $\lambda$, in randomly 
generated (unit) directions. The steps involved in the Hit-and-Run sampler are 
detailed in Algorithm \ref{alg:har}. We employ this approach  since it ensures 
that the generated datapoints cover the feasibility region in a reasonably uniform 
manner \citep{METE20126, Zabinsky2013}. This in turn ensures that the network has 
observed training samples that span the entire feasible set on an average, thereby 
aiding its ability to generalize. 

\begin{algorithm}
\label{alg1}
\caption{Hit-and-Run Sampler}\label{alg:har}
\begin{algorithmic}[1]
\Procedure{Hit-and-Run }{$\mathcal{C}_{\infty}, N_S$}
\State Pick random point $\mathbf{x} \in \mathcal{C}_{\infty} = \{\mathbf{x} \in \mathbb{R}^n \mid \mathbf{C}_x\mathbf{x} \leq \mathbf{d}_x\}$
\State $\mathcal{S} \leftarrow \{\mathbf{x}\}$
\For{$i = 1, \hdots, N_S-1$}
\State $\lambda_i \leftarrow \infty$
\State Generate random unit direction $\mathbf{d}_i$
\For {$(\mathbf{c},d)$ in $(\mathbf{C}_x,\mathbf{d}_x)$}
\State $\lambda \leftarrow \displaystyle\frac{d-\mathbf{c}\cdot \mathbf{x}}{\mathbf{c}\cdot\mathbf{d}_i}$
\If {$\lambda > 0$} \Comment{To ensure right direction}
\State $\lambda_i \leftarrow \text{min}(\lambda_i, \lambda)$
\EndIf
\EndFor
\State $\lambda_i \leftarrow \text{drawn from }\mathbb{U}[0,\lambda_i)$
\State $\mathbf{x} \leftarrow \mathbf{x} + \lambda_i\mathbf{d}_i$
\State $\mathcal{S} \leftarrow \mathcal{S} \cup \{\mathbf{x}\}$
\EndFor
\State \textbf{return} $\mathcal{S}$
\EndProcedure
\end{algorithmic}
\end{algorithm}

\subsection{Training}
Once the training and test datasets are generated, the corresponding training outputs 
and gradients are obtained from using CVXPY to solve the MPC problem.
We then use a supervised learning method to train the networks on a data set 
$\mathcal{D} = \{(\mathbf{x}_i,\mathbf{u}_i, \mathbf{u}_i')\}$ of input-output pairs. 
During the training, we learn for the network parameters by minimizing the loss 
function $\mathcal{L}_{tr}(\pmb\theta)$ defined in \eqref{eq:train_loss} with respect 
to $\pmb{\theta}$. 

In order to increase the training speed, we split the data into smaller subsets 
(mini-batches) and compute the the training loss function (\ref{eq:train_loss}) 
for each mini-batch. Each mini-batch consists of five training samples. We then use 
the gradient descent-based \textit{Adam} optimizer \citep{adam} to backpropagate 
the loss and update the parameters $\pmb{\theta}$ following each batch. Once all 
the mini-batches have been iterated over, one training epoch is completed. We 
train the networks until $\mathcal{L}_{tr}(\pmb\theta)$ is reduced to $0.01$ or 
for a maximum of 50000 training epochs. 


\section{Examples}
\label{sec:examples}
We now consider the application of the proposed concepts on a set of networks 
with different regularization constants $\gamma$, trained on MPC problems for 
a two-dimensional system. We evaluate the networks in terms of two performance metrics:

\begin{enumerate}
    \item The normalized mean square error (NMSE) which is defined as
    \begin{align*}
        \mathrm{NMSE}=\frac{\mathrm{E}\|\hat{\mathbf{u}}-\mathbf{u}\|_2^2}{\mathrm{E}\|\mathbf{u}\|_2^2},
    \end{align*}
    where $\mathrm{E}$ denotes the average over all samples in the training or test dataset.
    \item The normalized control cost $J$, defined as the objective in Equation 
    (\ref{eq:mpc}) normalized by $\mathbf{x}(0)^T\mathbf{x}(0)$:
    \begin{align*}
        J =  \frac{\mathbf{x}(N)^T\mathbf{Q}_N\mathbf{x}(N)+ \sum_{k=0}^{N-1}\left[\mathbf{x}(k)^T\mathbf{Q}\mathbf{x}(k)+\mathbf{u}(k)^T\mathbf{R}\mathbf{u}(k)\right]}{\mathbf{x}(0)^T\mathbf{x}(0)}
    \end{align*}
    where $\mathbf{x}(0)$ is the initial state of the trajectory.
\end{enumerate}
Both the metrics are evaluated on test data, previously unseen by the networks during 
the training. The NMSE helps evaluate the control law predicted by the network with 
respect to the ground truth, whereas the control cost measures how well the control 
law is in terms of minimizing the control objective: the smaller the $J$, the better 
the control achieved.

\subsection{Two-Dimensional system}
\begin{figure}
    \centering
    \includegraphics[scale=0.4]{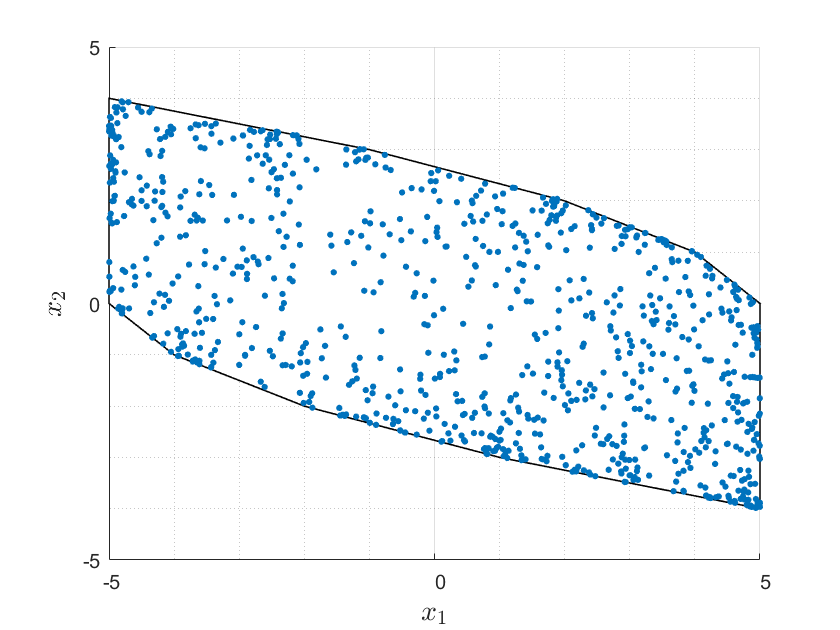}
    \caption{Hit-and-run sampling of 1000 points from the set $\mathcal{C}_{\infty}$ for system (\ref{eq:system}) with system and control matrices (\ref{eq:2D_system}) subject to the constraints (\ref{eq:2D_constraints}).}
    \label{fig:exA_har}
\end{figure}

We consider the example of a two-dimensional state vector with a scalar input under 
constraints. Despite being relatively low-dimensional, such a scenario occurs 
regularly in many real-life control applications. 
Let us then consider a two-dimensional system specified as follows \citep{mpc_book}:
\begin{equation}
    \mathbf{A} =
    \begin{bmatrix}
    1.0 & 1.0 \\
    0.0 & 1.0
    \end{bmatrix},
    \quad
    \mathbf{B} = 
    \begin{bmatrix}
    0.0 \\
    1.0
    \end{bmatrix}
    \label{eq:2D_system}
\end{equation}
subject to the constraints
\begin{equation}
    \begin{bmatrix}
    -5.0 \\
    -5.0
    \end{bmatrix}
    \leq
    \mathbf{x}(k)
    \leq
    \begin{bmatrix}
    5.0 \\
    5.0
    \end{bmatrix},
    \quad
    -2.0 \leq u(k) \leq 2.0,
    \quad k = 1, \hdots, N
    \label{eq:2D_constraints}
\end{equation}
and with cost parameters
\begin{equation}
    \mathbf{Q}_N = \mathbf{Q} = 
    \begin{bmatrix}
    1.0 & 0.0 \\
    0.0 & 1.0
    \end{bmatrix},
    \quad
    \mathbf{R} = 10,
    \quad
    N = 3.
    \label{eq:2D_costs}
\end{equation}
We generate training and testing data by first sampling states from $\mathcal{C}_{\infty}$ 
for the system (\ref{eq:2D_system}) subject to (\ref{eq:2D_constraints}) using the 
Hit and Run Algorithm 1, see Figure \ref{fig:exA_har} for an example of 1000 sampled points. 
We then solve for the optimal controls and corresponding gradients using CVXPY with cost 
parameters given by (\ref{eq:2D_costs}).

In our experimental setup we consider three sets of ten neural networks for four different 
settings of the regularization constant $\gamma \in \{0,\, 0.1,\, 1, \, 10\}$. For each set 
of networks we (pseudo) randomly generate 10 sets of training data, one for each network in 
the set. The first set of networks is trained using 25 samples, the second set using 50 
samples, and the third set using 100 samples. We use $S_{(25)}$, $S_{(50)}$ and $S_{(100)}$ 
to denote the three network sets.

For the NMSE evaluation, we use three separate testing datasets, one for each set of networks, 
all consisting of 100 samples. For the control cost evaluation, we sample sets of 100 initial 
states from $\mathcal{C}_\infty$ using Algorithm \ref{alg:har}. Starting from these, we then 
simulate the networks in closed loop for three time steps to generate trajectories.

\begin{figure}[h!]
    \centering
    \includegraphics[scale=0.5]{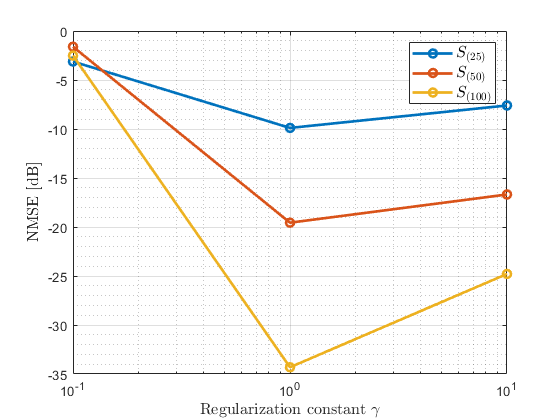}
    \caption{The test NMSE averaged over the ten networks in each network set as a function of the regularization constant $\gamma$.}
    \label{fig:gamma_0:1}
\end{figure}

In Figure \ref{fig:gamma_0:1} we plot the NMSE  averaged over the ten trained networks in 
each network set in the decibel [dB] scale. Table \ref{tab:nmse_tab_2d} compares the lowest 
overall NMSE for the cases $\gamma=0$ and $\gamma\neq0$. We see that $\gamma=1$ results in 
the lowest NMSE for each network set. This suggests that including gradient information 
during the training phase improves the performance of the network. From a theoretical 
point of view, the choice of $\gamma$ should not be so important since we do not have 
measurement noise. However, from a numerical point of view it makes a difference, for 
example in Figure \ref{fig:gamma_0:1} we note an increase in the NMSE for $\gamma > 10$. Notice that the we only evaluate the NMSE in the control mapping and not in its gradient.

\begin{figure}[h!]
    \centering
    \includegraphics[scale=0.5]{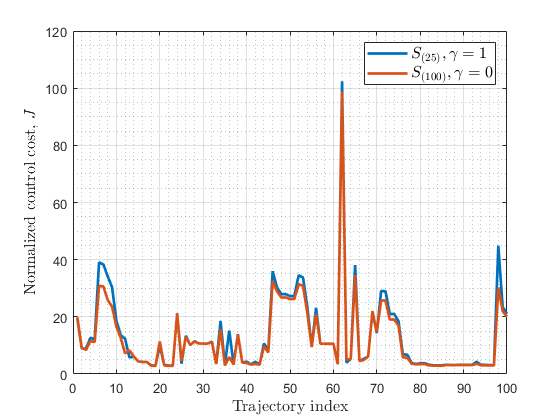}
    \caption{Control cost evaluation over 100 trajectories. The control costs are averaged over the ten networks in the respective sets $S_{(25)}$ and $S_{(100)}$.}
    \label{fig:control_cost_eval_3_simulsteps}
\end{figure}

Figure \ref{fig:control_cost_eval_3_simulsteps} shows a comparison of the the control 
costs for the two cases $\{S_{(25)}, \gamma=1\}$ and $\{S_{(100)}, \gamma=0\}$ averaged 
over the ten trained networks in each respective set. Table \ref{tab:control_cost_3step_2d} 
shows the control cost averaged over all trajectories. An interesting observation is that 
the performance of the setting $\{S_{(25)}, \gamma=1\}$ is comparable to that of the 
setting $\{S_{(100)}, \gamma=0\}$. In fact, a larger training dataset will in general 
lower the generalization error. The results then points to the richness of the gradient 
information.

Considering the proportions of the amount of training data to test trajectories, it is 
likely that there are clusters of samples (initial states) in regions that the network 
has not been exposed to during training, which might explain the peaks we observe in 
Figure \ref{fig:control_cost_eval_3_simulsteps}. A possible explanation is that the 
network in those cases produces either very large control signals and/or steers the 
state away from the reference state (possibly outside the feasible region), which 
would result in large control costs. \cite{chen2019large} suggest a projection 
strategy for ensuring feasibility, i.e. constraint satisfaction, of the generated 
control inputs by projecting them onto a safe region. The approaches in 
\citep{Winqvist1477561} is also based on this idea. We do not employ any such 
safety measures in our network structure. Note also that we do not explicitly 
train the networks to optimize the control cost, which is another possible approach 
for improving the results.  

\begin{table}
\centering
\begin{tabular}{ |p{1cm}|p{2.5cm}|p{2.5cm}|p{1cm}|  }
 \hline
 Set & {NMSE [dB] \newline (no regularization)} & NMSE [dB] \newline (regularization) & Best $\gamma$ \\
 \hline
 $S_{(25)}$   & 2.5285   & -9.8901  &   1\\
 $S_{(50)}$    & -0.8165  & -19.5597 &   1\\
 $S_{(100)}$   & 2.1549   & -34.3113 &   1\\
 \hline
\end{tabular}
\vskip0.5\baselineskip
\caption{NMSE evaluation for 2D example. The table presents the lowest NMSE when no regularization ($\gamma=0$) was used during training, as well as the lowest overall NMSE and the corresponding regularization constant $\gamma$. For all sets the lowest NMSE was found for $\gamma=1$.}
\label{tab:nmse_tab_2d}
\end{table}

\begin{table}
\centering
\begin{tabular}{ |p{2cm}|p{6cm}|}
 \hline
 $\{\text{Set}, \gamma \}$ & {$J$ (avg. over 100 test traj.)} \\
 \hline
 \{$S_{(25)}$, 1\}   & 13.3483  \\
 \{$S_{(100)}$, 0\}  & 12.2070  \\
 \hline
\end{tabular}
\vskip0.5\baselineskip
\caption{Control cost evaluation for 2D example.}
\label{tab:control_cost_3step_2d}
\end{table}


\section{Conclusion and Future Work}
\label{sec:conclusion}
We have presented a framework for off-line training and evaluation of a neural network 
approach for implementing MPC using gradient data. The underlying question is if it is 
possible to replace on-line MPC optimization solvers with trained NNs. This would allow 
for very efficient and robust real time implementations. At the same time, there is great 
progress in the area of embedded convex optimization for control. The  idea is to 
approximate the MPC mapping from state to control input with constrained ReLU based 
neural network. The main novel result is on how to include the gradient of the MPC 
controller with respect to the state input in the training of NNs. 
We have used CVXPY \citep{agrawal2019differentiable} and PyTorch \citep{NEURIPS2019_9015} 
to implement this framework. We also use CVXPY and cvxpylayers to generate training 
data and to evaluate the resulting controller. 

A key factor is the generation of samples for the off-line training, which related to 
input design in system identification, \citep{7879927}. This is a challenge when the 
dimension of the state space increase.  Here we proposed to use a hit-and-run sampler, 
and evaluated the resulting controller based on trajectories and  normalized cost-functions. 
The numerical tests showed the trade-off between  the number of training data and the 
approximation properties of the resulting controller.

This paper is a first step towards controller identification of MPC using ReLU networks. 
It should be noted that we do not assume any model information other than from $\mathbf{x}_i$, 
$\mathbf{u}_i$ and gradient in training the network-- so or approach is not restricted 
to eMPC or even time-invariant MPC problems. One can also use model parameters as inputs 
to the NN and  gradients with respect to them as training data. It is also possible to 
train the NN  to predict the $\mathbf{u}(k)$, $k=0,\ldots,N-1$, for the entire MPC 
horizon by giving just 
$\mathbf{x}(0)$ as input. In this way we are feeding the information that we need 
the control to take $N$ steps.

The next step is also to evaluate this approach on more advanced MPC control problems 
including nonlinear MPC. The numerical example studied here is for proof of concept. 
We also use very basic methods for evaluation. The approach would also benefit from 
training on trajectories instead of just on the control mapping. We will be pursuing 
these aspects in the future.

\bibliography{ifacconf}             
                                                   






\end{document}